\documentclass[aps,prl,floatfix,twocolumn,showpacs,10pt]{revtex4-1}

\usepackage{epsfig,amsmath,amssymb}

\begin{document}

\title{Theory of Spin Relaxation in Two-Electron Lateral Coupled Quantum Dots}
\author{Martin Raith$^1$, Peter Stano$^2$, Fabio Baruffa$^{1,3}$ and Jaroslav Fabian$^1$}
\affiliation{$^1$Institute for Theoretical Physics, University of Regensburg, D-93040 Regensburg, Germany\\
$^2$Institute of Physics, Slovak Academy of Sciences, 845 11 Bratislava, Slovakia\\
$^3$German Research School for Simulation Sciences, Forschungszentrum Juelich, D-52425  Germany}

\vskip1.5truecm

\begin{abstract}
A global quantitative picture of the phonon-induced two-electron spin relaxation in GaAs double quantum dots is presented using highly accurate numerics. Wide regimes of interdot coupling, magnetic field magnitude and orientation, and detuning are explored in the presence of a nuclear bath. Most important, the giant magnetic anisotropy of the singlet-triplet relaxation can be controlled by detuning switching the principal anisotropy axes: a protected state becomes unprotected upon detuning, and vice versa. It is also established that nuclear spins can dominate spin relaxation for unpolarized triplets even at high magnetic fields, contrary to common belief.
\end{abstract}


\maketitle


Electron spins in quantum dots \cite{loss1998:PRA} are among perspective candidates
for a controllable quantum coherent system in spintronics \cite{zutic2004:RMP, fabian2007:APS}.
Spin qubits in GaAs quantum dots, the current state of the art \cite{hanson2007:RMP,PhysRevLett.107.146801}, are coupled
to two main environment baths: nuclear spins, and phonons \cite{khaetskii2001:PRB}.
The nuclei dominate decoherence, which is on $\mu$s timescales. But only phonons are an efficient energy sink for the relaxation of the energy resolved spin states, leading to spin lifetimes as long as seconds \cite{amasha2008:PRL}.

The extraordinary low relaxation is boosted by orders of magnitude at spectral crossings, unless special conditions---such geometries we call easy passages---are
met \cite{golovach2004:PRL,stano2006:PRL}. Spectral crossings seem inevitable in the manipulation based on the Pauli spin blockade \cite{loss1998:PRA, hu2000:PRA}, the current choice in spin qubit experiments \cite{taylor2007:PRB}. On the other hand, a fast spin relaxation channel may be desired, e.g., in the dynamical nuclear polarization \cite{pfund2007:PRL, rudner2007:PRL, rudner2007b:PRL}.

The single-electron spin relaxation is well understood \cite{elzerman2004:N, stano2006:PRB}: it proceeds through acoustic phonons, in proportion to their density of states, which increases with the transferred energy. The matrix element of the phonon electric field between spin opposite states is nonzero due to spin-orbit coupling or nuclear spins. At anticrossings the matrix element is enhanced by orders of magnitude, even though the anticrossing gap is minute ($\sim\mu$eV). The relaxation rate can be either enhanced or suppressed, depending on whether the energy or the matrix element effects dominate.

The two electron relaxation rates were measured in single \cite{fujisawa2002:N, sasaki2005:PRL, meunier2007:PRL} and in
double \cite{petta2005:PRB, koppens2005:S, johnson2005:N} dots. Theoretical works so far mostly focused on single dots \cite{climente2007:PRB, golovach2008:PRB},
or vertical double dots \cite{chaney2007:PRB, shen2007:PRB}, in which the symmetry of the confinement potential lowers the numerical demands. A slightly deformed dot was considered in Refs.~\cite{florescu2006:PRB,olendski2007:PRB},
and a lateral coupled double dot in silicon in Ref.~\cite{wang2011:JAP}. What is key for spin qubit manipulation and most relevant for ongoing experiments, is the case of weakly coupled and biased coupled dots. In addition, the relative roles of the spin-orbit and hyperfine interactions in the spin relaxation in GaAs quantum dots has not yet been established.

The analysis of the two electron double dot relaxation is challenging because many parameters need to be considered simultaneously: the magnitude and orientation of the magnetic field, the orientation of the dot with respect to the crystallographic axes, the strength of the interdot coupling (parametrized by either tunneling or exchange energy) and the bias applied across the double dot (detuning). Here we cover {\it all} these parameters, \emph{including the nuclear bath}, providing specific relevant predictions for experimental setups \cite{footnote0}. Perhaps the most striking results are the existence of islands of inhibited spin relaxation in the magnetic field and detuning maps, and the switch of the two principal $C_{2v}$ axes along which the relaxation shows a minimum or maximum, as detuning is turned on. While singlets and polarized triplets relax by spin-orbit coupling, the spin-unpolarized triplet relaxation is dominated by nuclear spins over a wide parameter range (the spin-orbit induced anisotropy is wiped out), contrary to common belief. The predicted giant spin relaxation anisotropy is a unique and experimentally testable signature of spin-orbit spin relaxation, which can also be useful for spin nanodevices, as we argue in this paper.

\section{Model}

We consider a laterally coupled, top-gated GaAs double quantum dot patterned in the plane perpendicular to $\hat z = \left[001\right]$. In the two-dimensional and envelope function approximation, the Hamiltonian reads
\begin{equation}\label{Model:Hamiltonian}
	H = \sum_{i=1,2} \left(T_i + V_i + H_{Z,i} + H_{\text{so},i} + H_{{\rm nuc},i}  \right) + H_C ,
\end{equation}
where $i$ labels electrons. The single-electron terms are
\begin{eqnarray}\label{Model:HamiltonianTerms}
	T &=& \mathbf{P}^{2}/2m = \left(-\text{i} \hbar \nabla + e \mathbf{A}\right)^2/2m, \\
	V &=& \left(1/2\right) m \omega_{0}^2 \text{min}\{\left(\mathbf{r}-\mathbf{d}\right)^2 \! , \left(\mathbf{r}+\mathbf{d}\right)^2\} + e\mathbf{E}\cdot\mathbf{r}, \\
	H_Z &=& \left(g/2\right) \mu_{\text{B}} \boldsymbol{\sigma} \!\cdot\! \mathbf{B} , \\
	H_{\text{so}} &=& H_{\text{br}} + H_{\text{d}} + H_{\text{d3}},
\end{eqnarray}
the kinetic energy, the biquadratic confinement potential, the Zeeman term, and the spin-orbit couplings, respectively. The position and momentum vectors are two-dimensional, where $\hat x = \left[100\right]$ and $\hat y = \left[010\right]$. The proton charge is $e$ and the effective electron mass is $m$. The confinement energy, $E_0 = \hbar \omega_{0}$, and the confinement length, $l_{0} = \left( \hbar / m \omega_{0} \right)^{1/2}$, define the characteristic scales. The potential is minimal at $\pm\mathbf{d}$ and we call $2d/l_0$ the interdot distance. The electric field $\mathbf{E}$ is applied along the dot main axis $\mathbf{d}$. Turning on $\mathbf{E}$ shifts the potential minima relative to each other by the detuning energy $\epsilon=2eEd$. The magnetic field is $\mathbf{B} = \left( B_x, B_y, B_z \right)$. We use the symmetric gauge, $\mathbf{A} = B_z \left(-y,x\right)/2$, and $\boldsymbol{\sigma} = \left( \sigma_x, \sigma_y, \sigma_z \right)$ are the Pauli matrices. The Land\'{e} factor is $g$ and the Bohr magneton is $\mu_{\text{B}}$. The Bychkov-Rashba, and the linear and cubic Dresselhaus Hamiltonian read
\begin{eqnarray}\label{Model:HamiltonianHso}
	H_{\text{br}} &=& \left(\hbar/2ml_{\text{br}}\right) \left(\sigma_x P_y - \sigma_y P_x \right) , \\
	H_{\text{d}} &=& \left(\hbar/2ml_{\text{d}}\right) \left(-\sigma_x P_x + \sigma_y P_y \right) , \\
	H_{\text{d3}} &=& \left(\gamma_c/2\hbar^3\right) \left(\sigma_x P_x P_{y}^2 - \sigma_y P_yP_{x}^2 \right) + \text{H.c.} ,
\end{eqnarray}
parameterized by the spin-orbit lengths $l_{\text{br}}$ and $l_{\text{d}}$, and a bulk parameter $\gamma_c$. Nuclei, labeled by $n$, couple through
\begin{equation}
H_{\rm nuc} = \beta \sum_n  {\bf I}_n \cdot \boldsymbol{\sigma}\, \delta ({\bf r} - {\bf R}_n),
\end{equation}
where $\beta$ is a constant, and ${\bf I}_n$ is the spin of a nucleus at the position ${\bf R}_n$. The Coulomb interaction is $H_C = e^2/4\pi\epsilon \left|\mathbf{r}_1-\mathbf{r}_2\right|$, with the dielectric constant $\epsilon$. The Hamiltonian, Eq.~\eqref{Model:Hamiltonian}, and its energy spectrum is discussed in Refs.~\cite{baruffa2010:PRL,baruffa2010:PRB}, including our numerical method (configuration interaction) for its diagonalization. Here we extent it by including nuclear spins, which we treat by averaging over unpolarized random ensemble. See Supplementary \cite{supp0} for further details.

The relaxation is mediated by acoustic phonons
\begin{equation}\label{Model:HamiltonianHelph}
H_{\text{ep}} = \text{i} \sum_{\mathbf{Q}, \lambda} \sqrt{\dfrac{\hbar Q}{2 \rho V c_{\lambda}}} V_{\mathbf{Q},\lambda} \left[ b^{\dagger}_{\mathbf{Q},\lambda} \text{e}^{\text{i} \mathbf{Q} \cdot \mathbf{R}} \!-\! b_{\mathbf{Q},\lambda} \text{e}^{-\text{i} \mathbf{Q} \cdot \mathbf{R}} \right] \! ,
\end{equation}
with deformation, $V_{\mathbf{Q},l}^{\text{df}}=\sigma_e$, and piezoelectric potentials, $V_{\mathbf{Q},\lambda}^{\text{pz}}=-2\text{i}eh_{14}(q_x q_y \mathrm{\hat e}_{\mathbf{Q},z}^{\lambda} +
q_z q_x \mathrm{\hat e}_{\mathbf{Q},y}^{\lambda}+q_y q_z \mathrm{\hat e}_{\mathbf{Q},z}^{\lambda}) /Q^3$.
The phonon wave vector is $\mathbf{Q}$, and the electron position vector is $\mathbf{R} = \left( \mathbf{r}, z \right)$. The polarizations are given by $\lambda$, the polarization unit vector reads $\mathbf{\hat e}$, and the phonon annihilation (creation) operator is denoted by $b$ ($b^{\dagger}$). The mass density,  the volume of the crystal, and the sound velocities are given by $\rho$, $V$, and $c_{\lambda}$, respectively. The phonon potentials are parameterized by $\sigma_e$, and $h_{14}$.

We define the relaxation rate as the sum of the individual transition rates to all lower-lying states for both piezoelectric and deformation potentials. Each rate (from $|i\rangle$ to $|j\rangle$) is evaluated using Fermi's Golden Rule in the zero-temperature limit,
\begin{equation}\label{Model:rate}
\Gamma_{ij} = \dfrac{\pi}{\hbar \rho V} \sum_{\mathbf{Q}, \lambda} \dfrac{Q}{c_{\lambda}} \left| V_{\mathbf{Q},\lambda} \right|^2 \left| M_{ij} \right|^2 \delta(\omega_{ij} - \omega_{\mathbf{Q}}) ,
\end{equation}
where $M_{ij} = \langle i | \text{e}^{\text{i} \mathbf{Q} \cdot \mathbf{R}} | j \rangle$ is the matrix element of the states with energy difference $\hbar \omega_{ij}$. Here we are interested in the rates of the singlet ($S$) and the three triplets ($T_+,T_0,T_-$) at the bottom of the energy spectrum.

\begin{figure}
 \centering
 \includegraphics[width=0.99\linewidth]{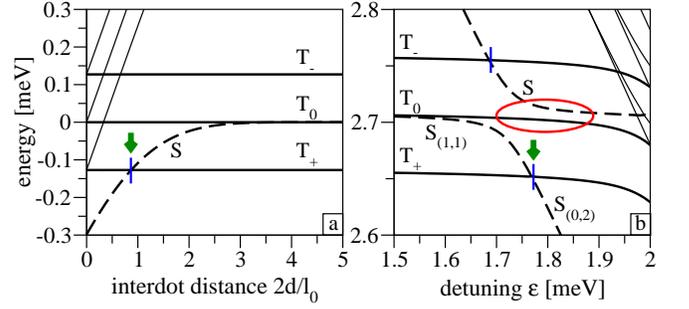}
\caption{Calculated energies of the lowest states for (a) variable interdot coupling (at $B=5$ T), and (b) detuning (at $B=2$ T). Singlet states are given by dashed, triplets by solid lines. The blue strokes mark singlet-triplet anticrossings. In (a), the energy of $T_0$ is subtracted, and in (b), the quadratic trend in $E$ is subtracted. The green arrows denote points of exact compensation and the red oval in (b) shows where nuclear spins dominate the $T_0$ relaxation.}
 \label{fig:1}
\end{figure}

In numerics we use GaAs parameters: $m=0.067m_e$, with $m_e$ the free electron mass, $g=-0.44$, $c_{l}=5290$ m/s, $c_{t}=2480$ m/s, $\rho=5300$ kg/$m^3$, $\sigma_e=7$ eV, $eh_{14}=1.4\times10^9$ eV/m, $\epsilon=12.9$, $\gamma_c=27.5$ eV\AA$^3$, $\beta=2\,\mu$eV nm$^3$, I=3/2. We choose typical lateral dots values, $l_{\text{br}}=2.42$ $\mu$m, $l_{\text{d}}=0.63$ $\mu$m, ${\bf d}$ $||$ [110] and the confinement energy $E_0 = 1.0$ meV, corresponding to $l_{0} = 34$ nm.

\section{Symmetric Double Dot}

We start with an unbiased double dot. We plot its spectrum in Fig.~\ref{fig:1}a) as a function of the interdot coupling, which translates into an exponentially sensitive $S-T_0$ exchange splitting $J$. Electrical control over $J$, necessary e.g.~to induce the $\sqrt{\text{SWAP}}$ gate \cite{loss1998:PRA}, allows for a fast switching between the strong and weak coupling regime, corresponding to the exchange splitting being larger and smaller than the Zeeman energy, respectively. During this switching, the ground state changes at an $S-T_+$ anticrossing.

\begin{figure}
 \centering
 \includegraphics[width=0.76\linewidth]{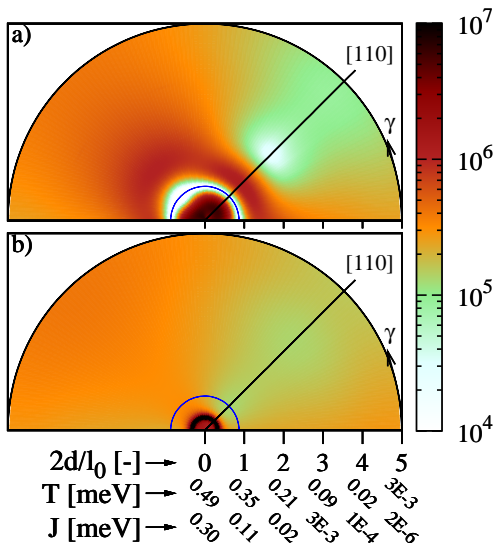}
\caption{Calculated relaxation rates of (a) the first excited state ($S$ or $T_+$, see Fig.~\ref{fig:1}a)) and (b) the triplet $T_0$ as a function of the in-plane magnetic field orientation $\gamma=\arccos(B_x/B)$ (angle) and the interdot distance $2d/l_0$ (radius of the polar plot), for a double dot at $B=$ 5 T. The x and y axes correspond to crystallographic axes $\left[100\right]$ and $\left[010\right]$, respectively. The dot orientation ${\bf d}$ $||$ $\left[110\right]$ is marked by a line. The blue half circles indicate the $S-T_+$ anticrossing, also marked on Fig.~\ref{fig:1}a). The x axis is converted to the tunneling energy $T$ and the exchange $J$, in addition to $2d/l_0$. The rate is given in inverse seconds by the color scale. The system obeys $C_{2v}$ symmetry, so point reflection would complete the graphs.}
 \label{fig:2}
\end{figure}

We cover the freedom of the interdot coupling in Fig.~\ref{fig:2}. Panel a) shows the relaxation of the first excited state [$S$ or $T_+$, see Fig. \ref{fig:1}a)]. First to note is the strong relaxation suppression at the $S-T_+$ anticrossing as the transferred energy becomes very small. Remarkably, the anticrossing does not influence the rate of $T_0$, plotted at panel b), at all (the peak close to $d=0$ is due to an anticrossing with a higher excited state). Even though the dominant channel, $T_0 \rightarrow T_+$, is strongly suppressed here, its reduction is exactly compensated by the elsewhere negligible $T_0 \rightarrow S$ channel. The exact compensation arises for the relaxation {\it into} a quasi-degenerate subspace (we denote such cases on Fig.~1 by green arrows) if
\begin{equation}
\label{eq:exact compensation}
\Delta E \ll {\rm min} \{ E , \hbar c_\lambda /l_0 \}.
\end{equation}
Here $E$ is the transition energy and $\Delta E$ is the energy width of subspace (the anticrossing gap). Equation \eqref{eq:exact compensation} states that the energy width $\Delta E$ is too small to be resolved by either phonons with energy $E$ or electron wave function scale $l_0$ \cite{supp0}. The relaxation then proceeds into the subspace rather then into its constituent states, so that any mixing of the states within the subspace is irrelevant.

Further to note on Fig.~2 is the anisotropy of relaxation, which reflects the anisotropy of the spin-orbit fields. In the weak coupling regime, the relaxation rates are minimal if the magnetic field orientation is parallel to the dot main axis, which results in an isle of strongly prolonged spin lifetimes. Note that this is in contrast to the biased dot (see below), and to the single-electron case, where the minimal in-plane magnetic field direction, the easy passage, of a $\mathbf{d} \parallel \left[110\right]$ double dot is perpendicular to $\mathbf{d}$ \cite{PhysRevB.83.195318,stano2006:PRL}.
The switch can be understood from the effective, spin-orbit induced, magnetic field \cite{stano2006:PRL} if written using the coordinates along the dot axes $x_d, y_d=(x\pm y)/\sqrt{2}$,
\begin{equation}\label{Model:Bso}
{\bf B}_{\rm so}={\bf B} \times \{ x_d (l_{br}^{-1} - l_d^{-1}) [1\overline{1}0] + y_d (l_{br}^{-1} + l_d^{-1}) [110] \}/\sqrt{2}.
\end{equation}
At the anticrossing, the mixing due to $x_d$ is by far dominant, so the minimum appears with ${\bf B}$ along [1$\overline{1}$0]. This $x_d$ dominance will be the case for a biased dot, too. On the other hand, in a single dot  $x_d$ and $y_d$ induce comparable mixing, and ${\bf B}_{so}$ becomes minimal if the larger term (the one with $y_d$) is eliminated. Weakly coupled unbiased dot is in this respect similar to a single dot as the two-electron transitions can be understood as flips of a particular electron located in a single dot. Since the direction for the rate minimum switches upon changing $d$, the system does not show an easy passage, that is a low relaxation rate from weak to strong coupling regime.

\begin{figure}
 \centering
 \includegraphics[width=0.76\linewidth]{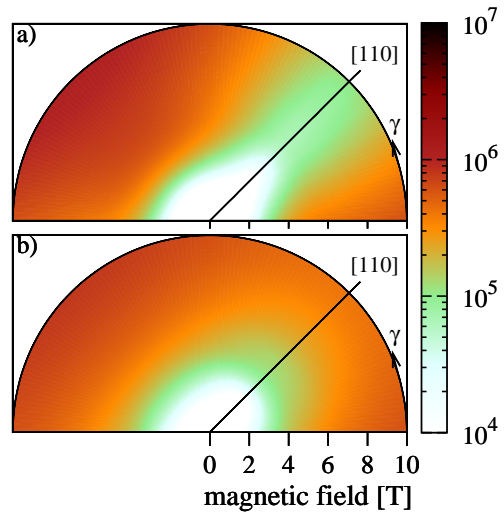}
\caption{Calculated relaxation rates of (a) the first excited state and (b) the triplet $T_0$ as a function of the in-plane magnetic field orientation $\gamma$ (angle) and the magnetic field magnitude (radius of the polar plot), for a double dot with $T$ = 0.1 meV. The layout with respect to the crystallographic axes is the same as in Fig.~2. The rate is given in inverse seconds by the color scale.}
 \label{fig:3}
\end{figure}

We plot the magnetic field dependence for a weakly coupled unbiased double dot in Fig.~\ref{fig:3} and observe similar behavior as in Fig.~2. The relaxation rate is minimal if ${\bf B}$ $||$ ${\bf d}$ throughout the shown parametric region. This is because the anticrossing and the related directional switch happens here at so small magnetic field that it is not visible at the figure resolution. For completeness, we note that the $T_-$ relaxation behavior is very similar to the one for $T_0$ on both Figs.~2 and 3, and we do not show it.

\section{Biased Double Dot}

We now consider a biased double dot. Its spectrum is shown in Fig.~\ref{fig:1}b) as a function of the detuning. The ground state singlet is in the (1,1) configuration (one electron in each dot) for low, and in the (0,2) configuration (both electrons in one dot) for large detunings. The crossover, a broad singlet-singlet anticrossing, is a key handle in spin measurement and manipulation \cite{taylor2007:PRB}. The low to large detuning crossover involves $S-T_\pm$ anticrossing, exploited for nuclear-spin pumping \cite{pfund2007:PRL,Reilly08082008}.

\begin{figure}
 \centering
 \includegraphics[width=0.76\linewidth]{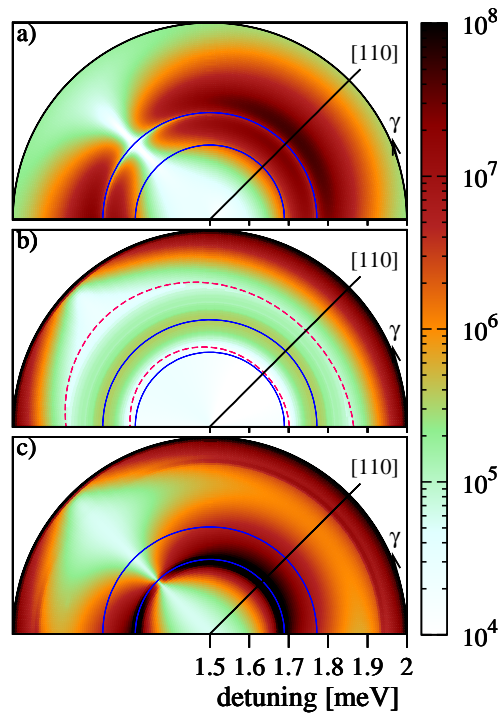}
\caption{Calculated relaxation rates of (a) the first excited state, (b) $T_0$, and (c) $T_-$ as a function of the in-plane magnetic field orientation $\gamma$ (angle) and detuning energy (radius of the polar plot), for a double dot with $2d/l_0=4.35$ ($T$ = 10 $\mu$eV), chosen along Ref.~\cite{taylor2007:PRB}, and $B=$ 2 T. The layout with respect to the crystallographic axes is the same as in Fig.~2. The rate is given in inverse seconds by the color scale. The blue lines indicate the singlet-triplet anticrossings, in line with the marks in Fig.~\ref{fig:1}b). The dashed red lines in panel b) confine the area where hyperfine coupling dominates.}
 \label{fig:4}
\end{figure}

We show the detuning and magnetic field influence on the relaxation in Fig.~\ref{fig:4}. At the singlet-triplet anticrossings, we observe that first, the relaxation rate of the first excited state dips at the $S-T_+$ anticrossing (though the dip is very narrow and hard to see at the figure resolution), and second, the $T_-$ rate strongly peaks at the $S-T_-$ anticrossing. This is a demonstration of the dominant effect of the anticrossing on the transition energy, and matrix element, respectively. Third, there are no other manifestations of the $S-T_\pm$ anticrossings, a fact due to the exact compensation already mentioned before. The anisotropy features of this geometry are striking. In the given range of detuning energies, states except $T_0$ exhibit a very distinctive easy passage for a magnetic field along $\left[1\bar10\right]$, where the relaxation is up to to three orders of magnitude smaller than with ${\bf B}$ along $\left[110\right]$. Though the directional switch occurs---rates become minimal for a magnetic field along $\left[110\right]$, it is again out of the figure scope (very small and very large detunings). The rates increase at detunings $\gtrsim2$ meV, because of spectral crossings with excited triplets, Fig.~\ref{fig:1}b), regime normally avoided in experiments. Double dots, with their spectral idiosyncrasies, are a unique system to observe a giant amplification of the spin-orbit anisotropies by a physical observable with bias control.

In large parts of the parametric space the relaxation of $T_0$ is dominated by nuclear spins, thus being isotropic. This is surprising, since the effective (Overhauser) nuclear magnetic field $B_{\rm nuc}$ is of the order of mT, much smaller than the spin-orbit field in Eq.~\eqref{Model:Bso}, $B_{\rm so}\sim(l_0/l_{\rm so})B\approx 30$ mT at $B=1$ T for our parameters. One therefore expects the nuclei to lead to much slower relaxation than the spin-orbit coupling. This was indeed the case for the unbiased dots and Figs.~2 and 3. How can then nuclei dominate here? 
Looking on Fig.~\ref{fig:1}b, this happens when states $T_0$ and $S(1,1)$ are nearby in energy. Here, the otherwise negligible hyperfine effects take over, {\it because the spin-orbit induced mixing of these two states is forbidden} \cite{florescu2006:PRB}. Estimating the wave function admixture in the lowest order, the nuclei dominate if
\begin{equation}
B_{\rm so} / |E_{T_0}-E_k| \lesssim B_{\rm nuc} / |E_{T_0}-E_S|,  
\end{equation}
with $k$ being the closest state to which $T_0$ is coupled by the spin-orbit interaction. The above condition generalizes in an obvious way for other states than $T_0$ and there are additional cases of nuclear dominance in our system. However, they  happen on parameter regions too small to be visible on the resolution of Fig.~4, so we discuss them only in the Supplementary material \cite{supp0}.

\section{Conclusions}

Our predictions are experimentally observable. Until now the spin-orbit origin, and especially its induced
directional anisotropy of the spin relaxation in weakly coupled two-electron dots has not yet been experimentally established. With employing vector magnets it should now be possible to overcome earlier experimental challenges and
change the magnetic field orientation while keeping the sample fixed and detect the anisotropy \cite{footnote5}.
The spin-orbit/nuclear induced relaxation can be masked by  cotunneling and smeared by a finite temperature. The former is reduced in the charge sensing readout setups \cite{barthel2009:PRL}, in which the coupling to the leads can be made small. The latter effect is small for experimentally relevant sub Kelvin temperatures, such that the directional anisotropies are well preserved.

Our results demonstrate control over the spin-orbit induced anticrossing gaps (easy passages appear if the gaps are closed) by sample and magnetic field geometry. It offers electrical tunability of spin relaxation, by changing the double dot orientation (in the Supplementary material \cite{supp0}, we suggest a spin current measurement device exploiting easy passage). In addition, such control may be especially useful when dealing with hyperfine spins. Indeed, in the polarization scheme considered in Ref.~\cite{rudner2007b:PRL}, the nuclear spin polarization is proportional to non-hyperfine assisted spin relaxation (see Eq.~(7) therein) and so would benefit from a setup with maximized spin-orbit induced relaxation rates (out of the easy passage). On the other hand, the adiabatic pumping scheme demonstrated in Ref.~\cite{Reilly08082008}, relies on the $S$-$T_+$ anticrossing being solely due to the nuclear spins (and not the spin-orbit coupling), suggesting improved efficiency in an easy passage configuration. We propose a similar non-adiabatic nuclear pumping scheme based on the easy passage in the Supplementary material \cite{supp0}. All these examples illustrate the potential benefits which intentional control of spin relaxation, based on our results, may offer.

\section{Acknowledgments}

\acknowledgments
This work was supported by DFG under grant SPP 1285 and SFB 689. P.S. acknowledges support by meta-QUTE ITMS NFP 26240120022, CE SAS QUTE, EU Project Q-essence, APVV-0646-10 and SCIEX.

\section{Supplemental Information}

\appendix

In this supplementary material, we provide further details illustrating the main text, and derivations of some of its results. In Sec.~A we comment on our numerical method and derive the discretized form of the electron-nuclear Hamiltonian, Eq.~(9). In Sec.~B we derive Eq.~(12), the condition for the exact compensation and illustrate the exact compensation showing channel resolved relaxation rates. In Sec.~C we compare the nuclear vs.~spin-orbit induced relaxation rates and discuss the additional cases to the one mentioned in the main text, where the nuclear spins dominate the relaxation. Finally, in Sec.~D we suggest schemes for dynamic nuclear polarization and detection of spin polarization, which are based on easy passages.  

\section{Numerical method}

We use the exact diagonalization of Eq.~(1) in the configuration interaction method. In this work, the two-electron basis consists of 1156 Slater determinants, generated by 34 single electron orbital states. The discretization grid is typically $135\times135$. The relative error for energies is below $10^{-5}$. The diagonalization procedure was described in detail in Ref.~[32]. Here we extend it adding nuclear spins, for which we now derive the discretized form of the Hamiltonian in Eq.~(9), $H^{\rm disc}_{\rm nuc}$.

Consider a basic element of the spatial grid, a rectangular box with lateral dimensions $h_x$ and $h_y$. Such volume elements are labeled by the index $k=1,...,M$, with $M$ their amount. In the two dimensional approximation one assumes that the electron wave function along the $z$ direction is fixed to $\psi(z)$. Let it be, for concreteness, the ground state of a hard-wall confinement of width $w=11$ nm.
As follows from Eq.~(9), the discretized Hamiltonian is diagonal in the spatial index. It has matrix elements 
\begin{equation}
( H^{\rm disc}_{\rm nuc})_{kk^\prime} = \delta_{k k^\prime} \beta \sum_{n \in k} {\bf I}_n \cdot \boldsymbol{\sigma} \langle k,\psi | \delta({\bf r}-{\bf R}_n) | k,\psi \rangle,
\end{equation}
where the sum is over nuclei inside the volume element $k$. To proceed further, we discretize the delta function in two dimensions as $1/h_x h_y$, introduce the nuclei volume density as $1/v_0$, and get
\begin{equation}
( H^{\rm disc}_{\rm nuc})_{kk^\prime} = \delta_{k k^\prime} (\beta/v_0) \sum_{n \in k} v_0 {\bf I}_n \cdot \boldsymbol{\sigma} |\psi(z_n)|^2 /h_x h_y.
\end{equation}
We now replace the sum over (typically many) nuclear spins by an effective spin $\boldsymbol{\mathcal{I}}$ and get the discretized form of the Hamiltonian as
\begin{equation}
( H^{\rm disc}_{\rm nuc})_{kk^\prime} = \delta_{k k^\prime} (\beta/v_0) \boldsymbol{\mathcal{I}} \cdot \boldsymbol{\sigma}.
\end{equation}
By the central limit theorem, the effective spins are completely described by their average and dispersion, which follow from the corresponding characteristics of the nuclear spin ensemble. For random unpolarized nuclear spins, which we consider, it holds
\begin{equation}
\langle {\bf I}_n \rangle = 0, \qquad  \langle {\bf I}_n\cdot{\bf I}_m \rangle = \delta_{nm} I(I+1),
\end{equation}
so that the effective spins have zero average and the following dispersion
\begin{equation}
\langle \boldsymbol{\mathcal{I}}_k \cdot \boldsymbol{\mathcal{I}}_{k^\prime} \rangle = \delta_{k k^\prime} I(I+1) / N.
\end{equation}
Here $N$ is the number of nuclei in the grid volume element, $N=h_x h_y h_z / v_0$, where the effective extension along $z$ is defined by the wave function profile
\begin{equation}
h_z^{-1} = \int\, {\rm d}z |\psi(z)|^4.
\end{equation}
For the hard-wall potential one gets $h_z = 2w/3$.

To calculate relaxation rates, we average Eq.~(11) over typically 50 configurations of nuclear ensemble. A single such configuration is parametrized by a set of effective spins $\boldsymbol{\mathcal{I}}$ drawn from a random Gaussian ensemble described by Eq.~(A5), for which we diagonalize the two-electron Hamiltonian. Having the two-electron spectrum allows us to calculate the matrix elements $M_{ij}$ and energy differences $\omega_{ij}$ which enter the relaxation rates $\Gamma_{ij}$ in Eq.~(11).

\section{Exact compensation of relaxation rate channels}

In general, the relaxation rate channels significantly change at spectral anticrossings because of the strong mixing of states. We consider the total relaxation rate by summing over the individual relaxation channels of all lower lying states. Therefore, a change in one relaxation channel may be compensated by another channel, such that the total relaxation rate is smooth (no peak or dip) across the anticrossing. This generally happens if a state relaxes {\it into} a quasi-degenerate subspace of anticrossing states. We exemplify such a situation on Fig.~1b) by a green arrow and consider relaxation of $T_0$ (state $i$) at this detuning. We write the total relaxation of $T_0$ as
\begin{equation}\label{Model:rate1}
\Gamma_i = \sum_j \dfrac{\pi}{\hbar \rho V} \sum_{\mathbf{Q}, \lambda} \dfrac{Q}{c_{\lambda}} \left| V_{\mathbf{Q},\lambda} \right|^2 \left| M_{ij} \right|^2 \delta(\omega_{ij} - \omega_{\mathbf{Q}}) ,
\end{equation}
where the sum includes the two quasi-degenerate states $j_1=T_+$ and $j_2=S$. The exact compensation arises if one can approximate the argument of the delta function by a common energy difference $\omega_{ij_1} \approx \omega_{ij_2} \approx \omega$. Indeed, the relaxation can be then written as
\begin{equation}\label{Model:rate2}
\Gamma_i = \dfrac{\pi}{\hbar \rho V} \sum_{\mathbf{Q}, \lambda} \dfrac{Q}{c_{\lambda}} \left| V_{\mathbf{Q},\lambda} \right|^2 \langle i | M P M^\dagger |i \rangle \delta(\omega - \omega_{\mathbf{Q}}),
\end{equation}
where $P_j=\sum_j |j\rangle \langle j |$ is the projector on the quasi-degenerate subspace, which is not influenced by mixing of the basis states $j$. The condition for the approximation Eq.~(B2) to be valid is that both under-integral factors in Eq.~(B1), the phonon density of states as well as electron overlap integral, are not changed much by the slight shift of the transition energy. The phonon density of states scales as a certain power (albeit different for piezoelectric and deformation potential) in the phonon wave vector, which translates into the condition
\begin{equation}
\omega_{j_1j_2} / c_\lambda \ll \omega_{ij} / c_\lambda.
\end{equation}
On the other hand, the natural scale for the electron wave function is the confinement length $l_0$, so that the overlap $M$ will not change much if 
\begin{equation}
\omega_{j_1 j_2} / c_\lambda \ll 1/l_0.
\end{equation}
Denoting $\omega_{j_1 j_2} = \Delta E / \hbar$ we get Eq.~(12) of the main text. For the interdot coupling denoted on Fig.1a and the dominant piezoelectric phonons we have $E=125$ $\mu$eV, $\hbar c_t / l_0=48\,\mu \text{eV}$, while $\Delta E$ is just $7\,\mu \text{eV}$, so that the exact compensation condition is well satisfied.

\begin{figure}
 \centering
 \includegraphics[width=0.98\linewidth]{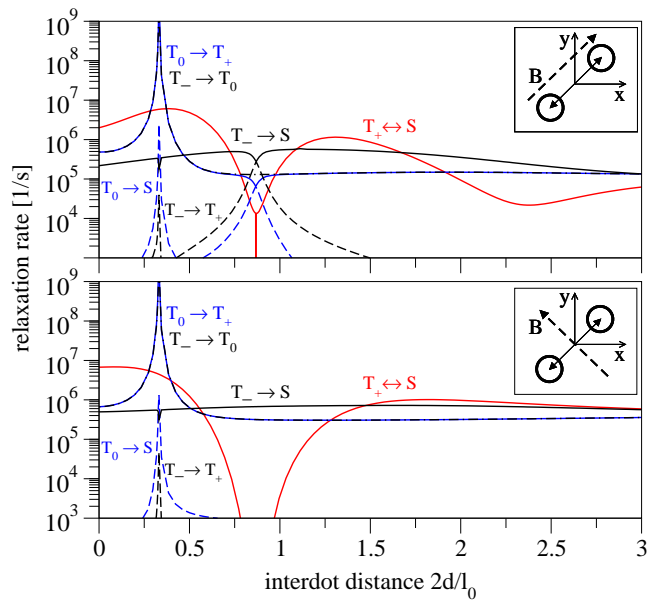}
\caption{Calculated channel resolved relaxation rates vs.~interdot distance in units of $l_0$ for both parallel (top) and perpendicular to $\mathbf{d}$ (bottom) in-plane magnetic field orientation ($B=5$ T, zero detuning). The relaxation channels of $T_0$ and $T_-$ are in blue and black color, respectively. The relaxation rate of the first excited state is red.}
 \label{fig:relaxation_channels}
\end{figure}

To illustrate the exact compensation, we plot in Fig.~\ref{fig:relaxation_channels} the individual relaxation channels as a function of interdot distance. The parameters are chosen the same as in Fig.~2 of the main text. We find the exact compensation at the $S-T_+$ anticrossing for the $T_0$ and the $T_-$ relaxation. In the case of the unpolarized triplet, the dip of the $T_0 \rightarrow T_+$ channel is compensated by a peak of the $T_0 \rightarrow S$ channel. For $T_-$, the dip and peak occurs in the $T_- \rightarrow S$ and $T_- \rightarrow T_+$ channels, respectively. Note that if the in-plane magnetic field is perpendicular to the dot main axis $\mathbf{d}$ (lower panel), the relaxation channels for $T_-$ and $T_0$ do not vary at all, as the $S-T_+$ anticrossing gap vanishes, $\Delta E=0$, and the exact compensation is trivial.

\section{Hyperfine versus spin-orbit induced relaxation}

\begin{figure}
 \centering
 \includegraphics[width=0.98\linewidth]{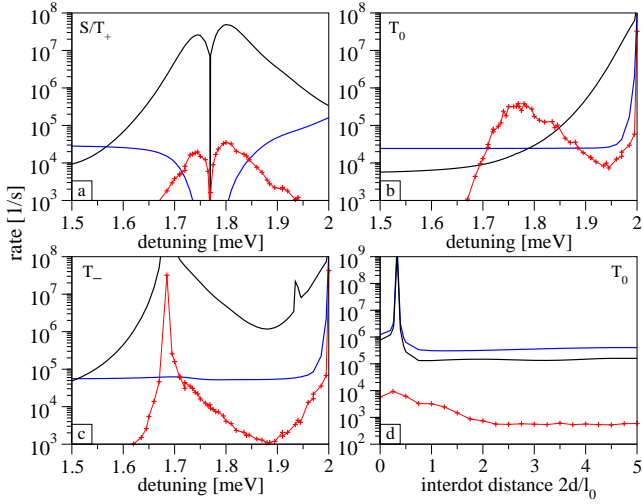}
\caption{Calculated spin-orbit induced relaxation rates for an in-plane magnetic field orientation parallel (black curves) and perpendicular (blue curves) to the dot main axis $\mathbf{d}$. The red curves show the hyperfine-induced spin relaxation. (a)-(c) Weakly coupled double dot ($T$ = 10 $\mu$eV) as a function of detuning for $B=2$ T. The panels display the relaxation rates for the first excited state, the unpolarized triplet, and $T_-$ respectively. (d) Unbiased double dot as a function of interdot distance (in units of $l_0$) for $B=5$ T. The relaxation rate of $T_0$ is shown.}
 \label{fig:hyperfine_relaxation}
\end{figure}

Comparing the value of the nuclear and spin-orbit effective fields, we estimate the relaxation due to the former is typically three orders of magnitude smaller, $(B_{\rm nuc}/B_{\rm so})^2 \sim 10^{-3}$, if the external field is of the order of Tesla. However, in a weakly coupled detuned double dot the nuclear spins can dominate over the spin-orbit induced relaxation in some cases, when Eq.~(14) is satisfied. We plot in Fig.~\ref{fig:hyperfine_relaxation} the spin relaxation rates enabled by spin-orbit and hyperfine coupling, respectively. Panel a) gives the relaxation rate of the first excited state. The hyperfine coupling becomes relevant only close to the $S-T_+$ anticrossing along the easy passage. Here, the wide dip is narrowed (red versus the blue curve). However, the rate remains reasonably low, such that the easy passage survives. Adding the nuclear dominated area to Fig.~4 a) of the main text would barely be visible. Panel b) shows the rate of $T_0$. We find that the hyperfine-induced relaxation is dominant for any in-plane magnetic field orientation if the unpolarized triplet is close in energy to the first excited singlet, as shown in Fig.~4 b) of the main text. Panel c) displays the relaxation of $T_-$. At the $S-T_-$ anticrossing, the spin-orbit induced relaxation strongly peaks unless the in-plane magnetic field orientation is perpendicular to the dot main axis. At the anticrossing, also the hyperfine-induced rate is enhanced. Displacing the magnetic field from the easy passage, the spin-orbit rate quickly gains on magnitude, therefore the nuclear-dominated area on Fig.~4 c) would cover only a single point at its current resolution. In panel d) we show the relaxation rate for an unbiased dot. We choose $T_0$ as an example, the state which is most prone to have relaxation dominated by nuclear spins in the biased dot. Here, the relaxation due to the spin-orbit coupling is several orders of magnitude larger than due to the nuclei, for any orientation of the external field. We observe a similar difference in rates for other states in this setup as well.

\section{Easy passage exploitation examples}

\subsection{Dynamical nuclear spin polarization}

\begin{figure}
 \centering
 \includegraphics[width=0.98\linewidth]{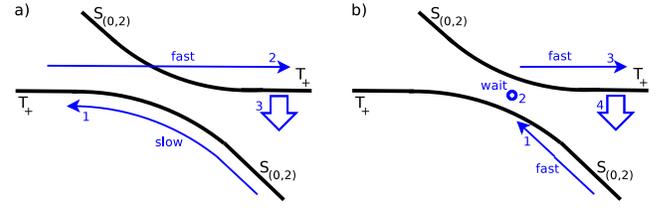}
\caption{Sketches of two schemes of dynamical nuclear spin pumping. The blue arrows indicate the path which the system state follows during one pump cycle.}
 \label{fig:DNSP}
\end{figure}

We sketch two schemes of dynamical nuclear spin pumping in Fig.~\ref{fig:DNSP}. The first is the one originally proposed by Reilly et.~al.~in Ref.~[35]. Here, the double dot is initialized in the $S_{(0,2)}$ state. Then the system is adiabatically brought through the anticrossing (step 1), by which a nuclear spin is flipped, assuming the anticrossing is due to the (transverse component of) the nuclear effective field (and not due to the spin-orbit coupling). Placing the system into the easy passage, which was not done in the experiment, thus offers improved scheme efficiency. The cycle is finished by resetting the system into the $S_{(0,2)}$, achieved by a fast transition (step 2) and a subsequent relaxation (step 3).

We propose here a non-adiabatic version of the scheme, which does not require the increasingly slower gap crossovers, by ending the step 1 inside the anticrossing. This necessarily requires to monitor the anticrossing position, which however is possible.
Again, the scheme is most efficient if the spin-orbit contribution to the anticrossing gap is minimized, what happens in the easy passage configuration.

\subsection{Spin polarization detection}

\begin{figure}
 \centering
 \includegraphics[width=0.75\linewidth]{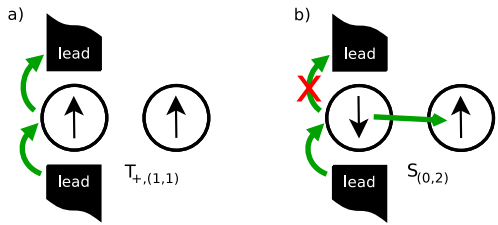}
\caption{Spin polarization detection scheme. (a) The double dot is in a $T_{+,(1,1)}$ state and the current is enabled. (b) For the $S_{(0,2)}$ state the current is blocked.}
 \label{fig:spin_current_measurement}
\end{figure}

Here we propose a device which allows to detect the spin polarization of a lead using a weakly coupled double dot. The dot is connected to source and drain leads such that the current passes only through the left dot. The system is biased such that only (1,1) and (0,2) occupations are allowed if the right dot is brought below in energy and the current is allowed to flow. The system is periodically brought above both source and drain leads so that the right dot is emptied. After such a reset, when the right dot is lowered in energy by a gate, the electron which is traversing the device may tunnel into the right dot and becomes trapped (we assume its spin is preserved). Now, if an electron with the same spin orientation enters the left dot, a $T_{+,(1,1)}$ triplet state is formed, with a long lifetime such that the electron in the left dot tunnels out and the current flows. If, on the other hand, a spin opposite electron enters the left dot, the system quickly collapses into the $S_{(0,2)}$ state and the current is blocked until the reset (see the sketch in Fig.~\ref{fig:spin_current_measurement}). As the result, the higher the spin polarization of the electrons in the source lead, the higher the current on average. The scheme requires 
\begin{equation}\label{append:condition}
\Gamma_{T_0 \to S} \gtrsim \Gamma_{\rm left \to lead} \gg \Gamma_{T_+ \to S}, 
\end{equation}
where $\Gamma$ are the rates for transitions corresponding to the indexes. The conditions in Eq.~\eqref{append:condition} are only achievable in the easy passage configuration while $\Gamma_{T_0 \to S}$ must be dominated by the hyperfine-induced relaxation.


\begin{thebibliography}{10}
\bibitem{footnote0}%
  \BibitemOpen
  \bibinfo {note} {We present the results for the state-of-the art GaAs quantum
  dots. We have also calculated the rates for silicon (unpublished), where the
  rates are orders of magnitude smaller but qualitatively the results are
  similar. Differences may arise for isotopically purified Silicon dots, where
  nuclei effects are reduced.}%
  \bibAnnoteFile{Stop}{footnote0}%
\end{thebibliography}
\end{document}